\documentclass[12pt]{iopart}

\usepackage{iopams}
\usepackage{graphicx}
\usepackage[toc,page,title,titletoc,header]{appendix}
\begin{document}
\bibliographystyle{unsrt}

\title{Virial
Theorem and Hypervirial Theorem in a spherical geometry}
\footnote{J. Phys. A: Math. Theor. 44 (2011) 365306 (11pp)}

\author{Yan Li$^{1}$, Fu-Lin Zhang$^{2}$$^{\dag}$, and Jing-Ling Chen$^{1}$$^{*}$}
\address{$^{1}$Theoretical Physics Division, Chern Institute of
Mathematics, Nankai University, Tianjin 300071, People's Republic of
China \\ $^{2}$Physics Department, School of Science, Tianjin
University, Tianjin 300072, People's Republic of China} \ead{{\dag}
flzhang@tju.edu.cn, {*} chenjl@nankai.edu.cn }

\begin{abstract}
The Virial Theorem in the one- and two-dimensional spherical
geometry are presented, in both classical and quantum mechanics.
Choosing a special class of Hypervirial operators, the quantum
Hypervirial relations in the spherical spaces are obtained. With the
aid of the Hellmann-Feynman Theorem, these relations can be used to
formulate a \emph{perturbation theorem without wave functions},
corresponding to the Hypervirial-Hellmann-Feynman Theorem
perturbation theorem of Euclidean geometry. The one-dimensional
harmonic oscillator and two-dimensional Coulomb system in the
spherical spaces are given as two sample examples to illustrate the
perturbation method.
%
%
%
%
%
%

\end{abstract}



\maketitle


\section {Introduction}\label{intro}

The Virial Theorem (VT) has been known for a long time in both
classical mechanics and quantum mechanics. In the classical case, it
provides a general equation relating the average over time of the
kinetic energy $\langle T\rangle$ with that of the function of
potential energy $\langle\vec{r}\cdot\nabla V\rangle$. The VT was
given its technical definition by Clausius in 1870
\cite{clausius1870xvi}. Mathematically, the theorem states
\begin{eqnarray}\label{0}
2\langle T\rangle=\langle\vec{r}\cdot\nabla V\rangle.
\end{eqnarray}
If the potential takes the power function $V(r)=\alpha r^n$ with
$r=|\vec{r}|$, the VT adopts a simple form as
\begin{eqnarray}\label{1}
2\langle T\rangle=n\langle V\rangle.
\end{eqnarray}
Thus, twice the average kinetic energy equals $n$ times the average
potential energy. The VT in quantum mechanics has the same form as
the classical one, except for the average over time in Eqs.
(\ref{0}) and (\ref{1}) replacing by the average over an energy
eigenstate of the system. It dates back to the old papers of Born,
Heisenberg and Jordan \cite{born1926quantenmechanik}, and is derived
from the fact that the expectation value of the time-independent
operator $\vec{r}\cdot\vec{p}$ under a eigenstate is a constant
\cite{schiff1968quantum},
\begin{eqnarray}\label{VTjiben}
i\frac{d}{dt}\langle\psi|\vec{r}\cdot\vec{p}|\psi\rangle=\langle\psi|[\vec{r}\cdot\vec{p},H]|\psi\rangle=0,
\end{eqnarray}
where $H=\frac{p^2}{2}+V$ is the Hamiltonian and $|\psi\rangle$ is
an eigenket of $H$.

In 1960, Hirschfelder \cite{hirschfelder1960classical} generalized
the relationship by pointing out that $\vec{r} \cdot \vec{p}$ could
be replaced by any other operators which were not dependent on time
explicitly. In this way, he established the Hypervirial Theorem
(HVT). For example, in a one-dimensional system, one can replace
$\vec{r}\cdot\vec{p}=x p$ by the \emph{hypervirial operator} $x^k
p$, and obtain the recurrence relation of $\langle x^k\rangle$,
\begin{eqnarray}\label{2}
2kE\langle x^{k-1}\rangle=2k\langle x^{k-1}V\rangle+\langle
x^k\frac{dV}{d x}\rangle-\frac{1}{4}k(k-1)(k-2)\langle
x^{k-3}\rangle,
\end{eqnarray}
where $k$ is an integer and $E$ is the eigenenergy.

 The Hellmann-Feynman (HF) Theorem is another important theorem in quantum mechanics, which has been
applied to the force concept in molecules by using  the internuclear
distance as a parameter
\cite{hellmann1937einf¨¹hrung,feynman1939forces}. Let the
Hamiltonian $H(\xi)$ of a system be a time-independent operator that
depends explicitly upon a continuous parameter $\xi$, and
$|\psi(\xi)\rangle$ be a normalized eigenfunction of $H(\xi)$ with
the eigenvalue $E_m(\xi)$, i.e.
$H(\xi)|\psi(\xi)\rangle=E_m(\xi)|\psi(\xi)\rangle$,
$\langle\psi(\xi)|\psi(\xi)\rangle=1$. The HF theorem states that
\begin{eqnarray}
\frac{\partial E_m(\xi)}{\partial
\xi}=\langle\psi(\xi)|\frac{\partial H(\xi)}{\partial
\xi}|\psi(\xi)\rangle.
\end{eqnarray}
If the potential takes the power function $V(r)=\alpha r^n$, the HF
gives an equation representing the relation between eigenenergy
$E_m$ and mean value of $r^n$,
\begin{eqnarray}\label{3}
\frac{\partial E_m}{\partial \alpha}=\langle r^n\rangle.
\end{eqnarray}

Based on the relations in Eqs. (\ref{2}) and (\ref{3}), the
Hypervirial-Hellmann-Feynman Theorem (HVHF) perturbation theorem is
established
\cite{swenson1972hypervirial,killingbeck1978perturbation}. It
provides  a very efficient algorithm for the generation of
perturbation expansions to large order,  replacing  the  formal
manipulation of Fourier series expansions  with  recursion
relations. This perturbation method just need the energy instead of
the wave functions of the system, and it is easy to achieve on the
computer.

\begin{figure}
\includegraphics[width=14cm]{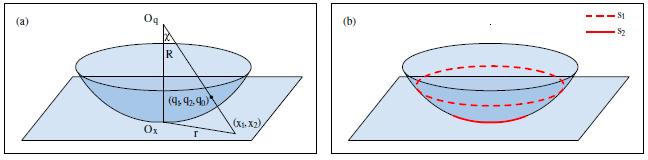} \\
 \caption{(a) The gnomonic projection,
which is the projection onto the tangent plane from the center of
the sphere in the embedding space. (b) Two classical orbits are
shown as $s_1$ (dashed line) and $s_2$ (solid line),  corresponding
to the circular motion and the radial one respectively.}\label{fig}
\end{figure}

These results are well known, but have not, to our knowledge, been
exploited in a curved space. In the present work, we focus on the
one- and two-dimensional spherical geometry. The coordinate systems
adopted in this paper are shown in Fig. \ref{fig} (a):
(\romannumeral1) An intuitive way to describe a two-dimensional
sphere is to embed it in a three-dimensional Euclidean space. Each
pair of independent variables $(q_1, q_2)$ of the three-dimensional
Cartesian coordinates $(q_1, q_2,q_0)$, with the origin $O_q$ in the
figure, under the constraint
\begin{eqnarray}
q_0^1+q_1^2+q_2^2=\frac{1}{\lambda}
\end{eqnarray}
corresponds to two points of the sphere, where $\lambda$ is the
curvature of the sphere. The points on the sphere can also be
described by the spherical polar coordinate $(R, \chi,\theta)$
defined by $(q_1, q_2,q_0)=(R\sin\chi\cos\theta,
R\sin\chi\sin\theta, R\cos\chi)$ with $R=1/\sqrt{\lambda}$ being a
constant. (\romannumeral2) The Cartesian coordinates $(x_1, x_2)$ of
the two-dimensional gnomonic projection,  which is the projection
onto the tangent plane from the center of the sphere in the
embedding space, is given by
\begin{eqnarray}
q_1=\frac{x_1}{\sqrt{1+\lambda r^2}},\ \ \ \ \ \ \ \ \
q_2=\frac{x_2}{\sqrt{1+\lambda r^2}}
\end{eqnarray}
where $r^2=x_1^2+x_2^2$ and the point of tangency $O_x$ in the
figure being the origin. And the polar coordinate $(r,\theta)$ of
the projection is defined by $r=R\tan\chi$ and $(x_1,
x_2)=(r\cos\theta, r\sin\theta)$. In this work, we mainly adopt the
two coordinate systems,
 $(x_1, x_2)$ and $(r,\theta)$, considering the
 results of Higgs  \cite{higgs1979dynamical} introduced in the
 following.


In 1979, Higgs \cite{higgs1979dynamical} introduced a generalization
of the hydrogen atom and harmonic oscillator in a spherical space.
He demonstrated that, in the gnomonic projection as shown in Fig.
\ref{fig} (a), the orbits of the motion on a sphere can be described
by
\begin{eqnarray}\label{guiji}
\frac{1}{2}L^2[r^{-4}(\frac{dr}{d\theta})^2+r^{-2}]+V(r)=E-\frac{1}{2}\lambda
L^2,
\end{eqnarray}
where the angular momentum $L=x_1p_2-x_2p_1$ is an invariant
quantity with the potential $V(r)$ being radial symmetric
The Hamiltonian can be written as
\begin{eqnarray}\label{Hamilton}
H=\frac{\pi^2}{2}+\frac{1}{2}\lambda L^2+V(r),
\end{eqnarray}
where
$\vec{\pi}=\vec{p}+\frac{\lambda}{2}\left[\vec{x}(\vec{x}\cdot\vec{p})+(\vec{p}\cdot\vec{x})\vec{x}\right]$
is the conserved vector in free particle motion on the sphere. Since
the curvature appears only in the right combination
$E-\frac{1}{2}\lambda L^2$ of Eq. (\ref{guiji}), the projected
orbits are the same, for a given $V(r)$, as in Euclidean geometry.
Consequently, according with the Bertrand Theorem
\cite{Bertrand,BertrandEnglish}, the orbits are closed only if the
potential takes the Coulomb or isotropic oscillator form, i.e.
$V(r)=-\frac{\kappa}{r}$ or $V(r)=\frac{1}{2}\omega^2r^2$, with
$\kappa$ and $\omega$ being constants. Therefore the systems
described by Eq. (\ref{Hamilton}) with the two mentioned potentials
are defined as the Kepler problem and isotropic oscillator in a
spherical geometry in \cite{higgs1979dynamical}. The algebraic
relations of their conserved quantities reveal the dynamical
symmetries of the two systems are described by the $SO(3)$ and
$SU(2)$ Lie groups respectively. These results are the beginning of
the so called \emph{Higgs Algebra}, which has been studied in a
variety of directions
\cite{bacry1966dynamical,KK,zhang2009higgs,JLChenHiggs,CS}.


The concept of symmetry is one of the cornerstones in the modern
physic, and dynamical symmetry plays a important role in many
important physical models. Since the dynamical symmetries of the
Kepler problem and isotropic oscillator on a $2$-sphere described by
Eq. (\ref{Hamilton}) adhere to the behaviors in two-dimensional
Euclidean geometry, our question is: Do there exist more qualities
of being homogeneous? This paper is aimed at constructing the VT and
the HVT for the spherical geometry and studying their applications.
In this work, we focus on the two- and one-demential cases for
simplicity. On the other hand, the motion on of a charged particle
on a $2$-sphere is not trivial, which is related with the famous
fractionally quantized Hall states
\cite{PhysRevB.83.115129,PhysRevLett.51.605}. We provide a general
equation relating the average of the kinetic energy with that of the
potential energy in the spherical geometry.
We also give a generalized HVHF, which could propose to solve a
class of problems the sense of perturbation.

 The article is
organized as follows: In the Sec. \ref{VT}, the VT in both classical
mechanics and quantum mechanics is constructed. In the Sec.
\ref{HVT}, we generalize the VT to HVT, and give the quantum
hypervirial relation. Two examples are taken to demonstrate the
perturbation method which is combined HVT with HF theorem in the
Sec. \ref{HFHV}. We end this paper with some relevant discussions in
the last section.

\section{Virial Theorem}\label{VT}


\subsection{Classical Mechanics}

To obtain the classical VT in the two-dimensional spherical
geometry, two we special orbits are listed in the following for
examples. (\romannumeral1) The first one is the uniform circular
motion with $\dot{r}=0$ as shown by the curve $s_1$ in Fig.
\ref{fig} (a). The kinetic energy of this case is given by
\[
T=\frac{1}{2}(R\sin\chi_0)^2 \dot{\theta}^2,
\]
where $R\sin\chi_0$ is the radius of the path. The corresponding
centripetal force is
\begin{eqnarray}\label{f1}
F=\frac{2T}{R\sin\chi_0}=\frac{(1+\lambda r^2)\vec{r} \cdot\nabla
V}{R\sin\chi_0}.
\end{eqnarray}
Hence, one can obtain
\begin{eqnarray}
2\langle T\rangle=\langle(1+\lambda r^2)\vec{r} \cdot\nabla
V\rangle,
\end{eqnarray}
which can be considered as the VT under the case of uniform circular
motion.
(\romannumeral2) The orbit $s_2$ in Fig. \ref{fig} (b) depictes the
case
 which the angular momentum $L$ is zero.  In the same way, the relationship between
kinetic energy and potential energy can be obtain as
\begin{eqnarray}\label{01}
2\langle(1+\lambda r^2)T\rangle=\langle(1+\lambda r^2)\vec{r}
\cdot\nabla V\rangle.
\end{eqnarray}
These serve a good inspiration for us to presume that the VT in a
spherical geometry is
\begin{eqnarray}\label{02}
2\langle(1+\lambda r^2)T_r\rangle+2\langle
T_\theta\rangle=\langle(1+\lambda r^2)\vec{r} \cdot\nabla V\rangle,
\end{eqnarray}
where $T_r$ and $T_\theta$ are the radial and rotational kinetic
energy.

In \ref{app}, we give a proof that Eq. (\ref{02}) is satisfied for
an arbitrary orbit in the spherical space, and it equivalents to
\begin{eqnarray}\label{jingdian}
\langle(1+\lambda r^2)\vec{r} \cdot\nabla V\rangle=\langle(1+\lambda
r^2)\pi^2\rangle=\langle(1+\lambda r^2)(2T-\lambda L^2)\rangle.
\end{eqnarray}
where $L=x_1p_2-x_2p_1$ is the angular momentum. 
It is easy to find that, when the curvature
$\lambda\rightarrow 0$, the above result reduces to Eq. (\ref{0}).

\subsection{Quantum Mechanics}
In the literature \cite{higgs1979dynamical}, to construct the the
conserved quantities on the sphere, Higgs replaced the momentum
$\vec{p}$ in the generators on the plan by the vector $\vec{\pi}$.
This enlightens us on the subject that we can replace
$\vec{r}\cdot\vec{p}$ in Eq.(\ref{VTjiben}) by
$\vec{r}\cdot\vec{\pi}+\vec{\pi}\cdot\vec{r}$ to obtain the VT on
the sphere. The expected value of the commutator is
\begin{eqnarray}\label{eq:1}
\langle[\vec{r}\cdot\vec{\pi}+\vec{\pi}\cdot\vec{r},H]\rangle=0.
\end{eqnarray}

For the system in the one-dimensional curve, whose Hamiltonian is
given by $H=\pi^2/2+V$ with $\pi=p+\lambda (x^2 p +p x^2)/2$, the
above relation leads to
\begin{equation}\label{eq:one-dimension}
\langle(1+\lambda x^2)\frac{\pi^2}{2}+\frac{\pi^2}{2}(1+\lambda
x^2)\rangle+\frac{1}{2}\langle\lambda(1+\lambda x^2)(1+3\lambda
x^2)\rangle=\langle(1+\lambda x^2)x\frac{dV}{dx}\rangle.
\end{equation}

And in the two-dimensional case, from Eq.(\ref{Hamilton}) and
Eq.(\ref{eq:1}), we obtain
\begin{equation}\label{eq:two-dimension} \langle(1+\lambda
r^2)\frac{\pi^2}{2}+\frac{\pi^2}{2}(1+\lambda
r^2)\rangle+\frac{1}{2}\langle\lambda(1+\lambda r^2)(2+3\lambda
r^2)\rangle=\langle(1+\lambda r^2)\vec{r}\cdot\nabla V\rangle.
\end{equation}
In the polar coordinate, the Hamiltonian (\ref{Hamilton}) can be
written as
\begin{eqnarray}\label{higgs}
&&H_0=T_r+T_\theta +V, \nonumber \\
 &&T_r=-\frac{1}{2}\biggr[3\lambda+\frac{15}{4}\lambda^2r^2+\frac{(1+\lambda
r^2)(1+5\lambda
 r^2)}{r}\frac{\partial}{\partial
 r}+(1+\lambda
 r^2)^2\frac{\partial^2}{\partial
 r^2}\biggr], \\
 &&T_\theta=-\frac{1}{2}\biggr[\frac{1}{r^2}\frac{\partial^2}{\partial\theta^2}+\lambda\frac{\partial^2}{\partial\theta^2}\biggr], \nonumber
\end{eqnarray}
where $T_r$ and $T_\theta$ denote the radial and rotational kinetic
energy. The relation of Eq. (\ref{eq:two-dimension}) equivalents to
\begin{equation}
\hspace{-1in}\langle(1+\lambda r^2)T_r+T_r(1+\lambda
r^2)\rangle+2\langle
T_\theta\rangle+\frac{1}{2}\langle\lambda(1+\lambda r^2)(2+3\lambda
r^2)\rangle=\langle(1+\lambda r^2)\vec{r}\cdot\nabla V\rangle.
\end{equation}
These results have the same form with the classical mechanical
counterparts in Eqs. (\ref{02}) and (\ref{jingdian}), but with a
term $\frac{1}{2}\langle\lambda(1+\lambda r^2)(2+3\lambda
r^2)\rangle$ in addition. And, the term is different from the
corresponding one in the one-dimensional case
(\ref{eq:one-dimension}), which comes from the commutation relation
of $\vec{r}$ and $ \vec{\pi}$.

\section{Hypervirial Theorems}\label{HVT}

In the above, we have got the VT in both classical mechanics and
quantum mechanics. We will discuss the quantum HVT in the present
part. A natural candidate of the hypervirial operator is $r^k\pi+\pi
r^k$ according to $r^k p$ in the plane we mentioned in Sec.
\ref{intro}, with $k$ being integers.

\subsection{One-dimensional}
In the one-dimensional case, one can calculate directly the
commutation relation in the expected value
\begin{equation}\label{eq:extended one-dimension}
\langle[x^k\pi+\pi x^k,H]\rangle=0,
\end{equation}
and obtain
\begin{eqnarray}
\hspace{-1in}[x^k\pi+\pi x^k,H]&=& i\frac{k}{2} \biggr\{2(1+\lambda
x^2)x^{k-1}\pi^2+2\pi^2(1+\lambda
x^2)x^{k-1} \nonumber\\
&&  +(1+\lambda x^2)\bigr[(k-1)(k-2)x^{k-3}+2\lambda
k^2x^{k-1}+\lambda^2(k+1)(k+2)x^{k+1}\bigr] \biggr\}\nonumber\\
&&-2i x^k(1+\lambda x^2)\frac{d V}{dx}.
\end{eqnarray}
Because of $\pi^2/2 = H-V$ and $\langle H\rangle = E_n$ being the
eigenvalues of the eigenstate, the Eq. (\ref{eq:extended
one-dimension}) turns to
\begin{eqnarray}\label{eq:one-dimension ditui}
\hspace{-0.5in}&&2kE_n\langle x^{k-1}\rangle_\lambda-2k\langle x^{k-1}V\rangle_\lambda-\langle x^k\frac{dV}{dx}\rangle_\lambda\nonumber\\
\hspace{-0.5in}&&+\frac{k}{4}\bigr[(k+1)(k+2)\lambda^2\langle
x^{k+1}\rangle_\lambda+2k^2\lambda\langle
x^{k-1}\rangle_\lambda+(k-1)(k-2)\langle
x^{k-3}\rangle_\lambda\bigr]=0,
\end{eqnarray}
in which we denote $\langle f \rangle_\lambda=\langle(1+\lambda
x^2)f \rangle$. Hence, we get the recurrence formula of $\langle
x^k\rangle_\lambda$, which is the  quantum  hypervirial relations in
the one-dimensional sphere.

\subsection{Two-dimensional}
We now consider HVT in the two-dimensional spherical geometry. For a
radial potential $V=V(r)$ in the Hamiltonian (\ref{higgs}), the
eigenfunction of energy can be written as
\begin{equation}
\Psi(r,\theta)=e^{im\theta}\psi(r),
\end{equation}
with $m=0, \pm1, \pm2...$ is the eigenvalue of the conserved angular
momentum $L$. The Schr\"{o}dinger equation
\begin{equation}
 H_0\Psi(r,\theta)=E\Psi(r,\theta)
\end{equation}
reduces to the radial equation as
\begin{equation}
H_1\psi(r)=E\psi(r),
\end{equation}
where the Hamiltonian $H_1$ is given by
\begin{equation}
\hspace{-1in}H_1=-\frac{1}{2}\left[(1+\lambda
r^2)^2\frac{d^2}{dr^2}+\frac{(1+\lambda r^2)(1+5\lambda
r^2)}{r}\frac{d}{dr}-\frac{1+\lambda r^2}{r^2}m^2+3\lambda
+\frac{15}{4}\lambda^2r^2\right]+V.
\end{equation}
It can be written as
\begin{eqnarray}\label{Hr}
H_1 =\frac{\pi_r^2}{2}+V_1,
\end{eqnarray}
where the radial component of $\vec{\pi}$ is $\pi_r=-i[(1+\lambda
r^2)\frac{d}{dr}+\frac{1}{2r}+\frac{3}{2}\lambda r]$ and
$V_1=V-\frac{1}{2}[(\frac{1}{2}-m^2)\lambda-\frac{m^2-1/4}{r^2}]$.
Choosing the hypervirial operator as $r^k\pi_r+\pi_rr^k$, one can
get the recurrence relation
\begin{eqnarray}\label{eq:two-dimension diui}
\hspace{-0.5in}&&2kE_n\langle r^{k-1}\rangle_\lambda-2k\langle r^{k-1}V_1\rangle_\lambda-\langle r^k\frac{dV_1}{dx}\rangle_\lambda\nonumber\\
\hspace{-0.5in}&&+\frac{k}{4}\left[(k+1)(k+2)\lambda^2\langle
r^{k+1}\rangle_\lambda+2k^2\lambda\langle
r^{k-1}\rangle_\lambda+(k-1)(k-2)\langle
r^{k-3}\rangle_\lambda\right]=0,
\end{eqnarray}
from
\begin{equation}
\langle[r^k\pi_r+\pi_rr^k,H_1]\rangle=0.
\end{equation}
Here, the notation $\langle f \rangle_\lambda=\langle(1+\lambda
r^2)f \rangle$. It is the two-dimensional quantum hypervirial
relation we will discuss in the present work. And when
$\lambda\rightarrow0$, it reduces to the result in the 2-plane case
\cite{XJZ3}.

\section{Application of The Hypervirial Theorems}\label{HFHV}
In this section, we will generalize the HVHF theorem to the
spherical space based on the hypervirial relations in the above.
When the perturbation of potential $V(r)$ takes the form as
$r^l(1+\lambda r^2)$ with $l$ being integers, we can determine the
eigenenergies in the various orders of approximation without
calculating the wavefunction, as the the HVHF theorem in the
Euclidean geometry. In the following, we will give two sample
examples to illustrate this method.
\subsection{One-dimensional Harmonic Oscillator}
The Hamiltonian of the one-dimensional harmonic oscillator in the
spherical geometry  with a perturbation potential is
\begin{eqnarray}
H=\frac{\pi^2}{2}+\frac{1}{2}\alpha x^2+\beta x^l(1+\lambda x^2),
\end{eqnarray}
where $\alpha, \beta$ are real numbers, $l$ is an integer and
$\lambda$ is the curvature of the sphere. The perturbation $\beta
x^l(1+\lambda x^2)$ has to be very small, and $\beta$ is the
smallness parameter.

Then, the HVHF recurrence relation in Eq. (\ref{eq:one-dimension
ditui}) becomes
\begin{eqnarray}\label{eq:ditui}
\hspace{-1in}\left[(k+1)\alpha-\frac{k}{4}(k+1)(k+2)\lambda^2\right]\langle
x^{k+1}\rangle_\lambda =&& 2kE_n\langle
x^{k-1}\rangle_\lambda+\frac{k^3}{2}\lambda\langle
x^{k-1}\rangle_\lambda+\frac{k}{4}(k-1)(k-2)\langle
x^{k-3}\rangle_\lambda \nonumber\\
\hspace{-1in}&&- \beta(2k+l)\langle x^{k+l-1}\rangle_\lambda
-\beta\lambda(2k+l+2)\langle x^{k+l+1}\rangle_\lambda. \ \ \
\end{eqnarray}
 The above equation establishes
 precisely regarding the $n$-th energy level. In order to obtain the approximate
 solution of the energy eigenvalues $E_n$, we expand both $E_n$ and desired expectation values $\langle x^k\rangle_\lambda$ in powers of the perturbation parameter
 $\beta$ as
 \begin{eqnarray}\label{eq:beta jishu e}
&&E_n=E_n^{(0)}+\beta E_n^{(1)}+\beta^2
E_n^{(2)}+\cdots=\sum_{j=0}^\infty\beta^j E_n^{(j)},\\
&&\langle x^k\rangle_\lambda=\langle
x^k\rangle_{\lambda,0}+\beta\langle
x^k\rangle_{\lambda,1}+\beta^2\langle
x^k\rangle_{\lambda,2}+\cdots=\sum_{j=0}^\infty\beta^j\mathcal
{Q}_j^k, \nonumber
\end{eqnarray}
where we introduce the notation $\mathcal {Q}_j^k=\langle
x^k\rangle_{\lambda,j}$ for convenience. We now insert the series in
(\ref{eq:beta jishu e}) into (\ref{eq:ditui}) and order in power of
$\beta$. It is straightforward to get the relation
\begin{eqnarray}\label{eq:q-ditui}
\hspace{-1in}&&\left[(k+1)\alpha-\frac{k}{4}(k+1)(k+2)\lambda^2\right]\mathcal
{Q}_\gamma^{k+1}\\
\hspace{-1in}&=&2k\sum_{j=0}^\gamma E_n^{j}\mathcal
{Q}_{\gamma-j}^{k-1}+\frac{k^3}{2}\lambda\mathcal
{Q}_\gamma^{k-1}+\frac{k}{4}(k-1)(k-2)\mathcal {Q}_\gamma^{k-3}
-(2k+l)\mathcal {Q}_{\gamma-1}^{k+l-1}-\lambda(2k+l+2)\mathcal
{Q}_{\gamma-1}^{k+l+1}.\nonumber
\end{eqnarray}

In addition, by the HF theorem, we know that
\begin{eqnarray}
\frac{\partial E_n}{\partial \beta}&=&\langle\frac{\partial
H}{\partial \beta}\rangle=\langle x^l\rangle_\lambda,
\end{eqnarray}
which gives another relationship of the coefficient of $\beta$:
\begin{eqnarray}\label{eq:H-F theory}
E_n^{(j)}=\frac{1}{j}\mathcal {Q}_{j-1}^l.
\end{eqnarray}
In other words, the $j$-th approximate of energy eigenvalue
$E_n^{(j)}$ is determined by the $(j-1)$-th approximate of desired
values $\mathcal {Q}_{j-1}^l$.

In the following, we would like to give an explicit example. We let
$l=1$ in the Eqs. (\ref{eq:q-ditui}) and (\ref{eq:H-F theory}) and
obtain, respectively,
\begin{eqnarray}
\label{eq:guanxi1}\hspace{-1in}&&\left[(k+1)\alpha-\frac{k}{4}(k+1)(k+2)\lambda^2\right]\mathcal {Q}_\gamma^{k+1}\\
\hspace{-1in}&=&2k\sum_{j=0}^\gamma E_n^{j}\mathcal
{Q}_{\gamma-j}^{k-1}+\frac{k^3}{2}\lambda\mathcal
{Q}_\gamma^{k-1}+\frac{k}{4}(k-1)(k-2)\mathcal {Q}_\gamma^{k-3}
-(2k+1)\mathcal {Q}_{\gamma-1}^{k} -\lambda(2k+3)\mathcal
{Q}_{\gamma-1}^{k+2},\nonumber
\end{eqnarray}
\begin{eqnarray}\label{eq:guanxi2}
E_n^{(j)}&=&\frac{1}{j}\mathcal {Q}_{j-1}^1.
\end{eqnarray}

One can start from
\begin{eqnarray}\label{eq:x0}
\langle x^0\rangle_\lambda=\langle 1+\lambda
x^2\rangle=1+\lambda\langle x^2\rangle
\end{eqnarray}
to obtain $\mathcal {Q}_j^0$. By the HF theorem
\begin{eqnarray}
\frac{\partial E_n}{\partial \alpha}&=&\langle\frac{\partial
H}{\partial \alpha}\rangle=\frac{1}{2}\langle x^2\rangle,
\end{eqnarray}
one can find that
\begin{eqnarray}
\frac{1}{2}\langle x^2\rangle=\sum_{j=0}\beta^j\frac{\partial
E_n^{(j)}}{\partial \alpha}.
\end{eqnarray}
Substituting it to Eq. (\ref{eq:x0}), the expectation value $\langle
x^0\rangle_\lambda$ expansion will be denoted as
\begin{eqnarray}\label{beta}
\langle x^0\rangle_\lambda=1+\lambda\langle x^2\rangle&=&\mathcal
{Q}_0^0+\beta\mathcal
{Q}_1^0+\beta^2\mathcal {Q}_2^0+\cdots\nonumber\\
&=&1+2\lambda\left[\frac{\partial E_n^{(0)}}{\partial
\alpha}+\beta\frac{\partial E_n^{(1)}}{\partial
\alpha}+\beta^2\frac{\partial E_n^{(2)}}{\partial
\alpha}+\cdots\right].
\end{eqnarray}
Ordering in power of $\beta$, it is easy to find the first term of
the recursion:
\begin{eqnarray}\label{eq:q-0-0}
\mathcal {Q}_0^0&=&1+2\lambda\frac{\partial E_n^{(0)}}{\partial
\alpha}=1+\frac{(2n+1)\lambda}{\sqrt{\lambda^2+4\alpha}}, \nonumber\\
\label{eq:q-1-0}\mathcal {Q}_1^0&=&2\lambda\frac{\partial
E_n^{(1)}}{\partial
\alpha},\\
\label{eq:q-2-0}\mathcal {Q}_2^0&=&2\lambda\frac{\partial
E_n^{(2)}}{\partial \alpha}, \nonumber\\
&\vdots&\nonumber
\end{eqnarray}
The eigenenergy of one-dimensional harmonic oscillator in a
spherical geometry is $E_{n}^{(0)}=(n+\frac{1}{2})
\frac{\lambda+\sqrt{\lambda^2+4\alpha}}{2}+\frac{n^2}{2}\lambda$
\cite{de1991operator,quesne2007spectrum}.

When $\gamma=0$, one can substitute $\mathcal {Q}_0^0$ into Eq.
(\ref{eq:guanxi1}) and obtain the values of $\mathcal {Q}_0^j$,
\begin{eqnarray}\label{eq:q-0-1}
&&\hspace{-1in}k=0\ \ \ \ \ \ \mathcal {Q}_0^1 =0,\\
&&\hspace{-1in}k=1\ \ \ \ \ \ \mathcal {Q}_0^2=\frac{[(2n+1)
\lambda+\sqrt{\lambda^2+4\alpha}]}{(4\alpha-3\lambda^2)\sqrt{\lambda^2+4\alpha}}\left[(2n+1)\sqrt{\lambda^2+4\alpha}+(2n^2+2n+3)\lambda\right],\\
\vdots\nonumber
\end{eqnarray}
Using Eq. (\ref{eq:q-0-1}) and (\ref{eq:guanxi2}), we can get the
first-order perturbation of $E_n$,
\begin{eqnarray}\label{eq_e_1}
E_n^{(1)}=\mathcal {Q}_0^1=0.
\end{eqnarray}
And from the Eq. (\ref{eq:q-1-0}) and (\ref{eq_e_1}), we have
\begin{eqnarray}
\mathcal {Q}_1^0=2\lambda\frac{\partial
E_n^{(1)}}{\partial\alpha}=0.
\end{eqnarray}
In the case of $\gamma=1$, using $\mathcal {Q}_1^0$ and Eq.
(\ref{eq:guanxi1}), we can derive the values of $\mathcal {Q}_1^j$,
and consequently the second approximation of energy level
\begin{eqnarray}
\hspace{-1in}E_n^{(2)}
&=&-\frac{\sqrt{\lambda^2+4\alpha}+(2n+1)\lambda}{2\alpha\sqrt{\lambda^2+4\alpha}}\nonumber\\
\hspace{-1in}&&-\frac{3\lambda[(2n+1)
\lambda+\sqrt{\lambda^2+4\alpha}]}{2\alpha(4\alpha-3\lambda^2)\sqrt{\lambda^2+4\alpha}}\left[(2n+1)\sqrt{\lambda^2+4\alpha}+(2n^2+2n+3)\lambda\right].
\end{eqnarray}
In this way, we can obtain the expectation value expansions
$\mathcal {Q}_\gamma^j$ and the energy values $E_n^{(j)}$ in the
various orders of approximation as
\begin{eqnarray}
\hspace{-1in}E_n^{(3)}&=&0,\\
\hspace{-1in}E_n^{(4)}&=&-\frac{1}{4\alpha}\left\{\left(\frac{6\lambda^2(2E_n^{(0)}+\lambda)}{4\alpha-3\lambda}+\lambda\right)\frac{\partial
\mathcal {Q}_1^1}{\partial \alpha}+2E_n^{(2)}\mathcal
{Q}_0^0-3\mathcal {Q}_1^1\right.\nonumber\\
\hspace{-1in}&&\ \ \ \ \ \ \ \ \
\left.-\frac{5\lambda}{3\alpha-6\lambda^2}\left[(4E_n^{(0)}+4\lambda)\mathcal
{Q}_1^1-(5+\frac{21\lambda E_n^{(0)}}{2\alpha-15\lambda^2})\mathcal
{Q}_0^2-\frac{21\lambda}{8\alpha-60\lambda}\mathcal
{Q}_0^0\right]\right\},\\
\hspace{-1in}&\vdots&\nonumber
\end{eqnarray}
In the limit $\lambda\rightarrow0$, $E_n^{(2)}$ is tending to
$-1/(2\alpha)$ and the other $E_n^{(j)}$ is tending to zero which
are corresponded with the exact result in the Euclidean space.

It is worth to mentioned that, alien from Euclidean space,
(\romannumeral1) The HF theorem has been used twice in this HVHF
perturbative method. (\romannumeral2) Only when the exponent $l$ in
the perturbation potential is a positive integer, we can get
$\mathcal
 {Q}_\gamma^j$ from Eqs. (\ref{eq:q-0-0}) (\ref{eq:guanxi1}) and
(\ref{eq:guanxi2}).

%

\subsection{Two-dimensional Coulomb System}
Here we wish  to show  that the HVHF  perturbation  method  can be
easily  applied  to treat the Coulomb system  with a perturbation in
the two-dimensional sphere which is described by the Hamiltonian
\begin{eqnarray}\label{hamilton}
H=\frac{\pi^2}{2}+\frac{1}{2}\lambda L^2-\frac{\kappa}{r}+\beta
r^l(1+\lambda r^2),
\end{eqnarray}
where $\kappa$ is a real number, and $\beta$ is the perturbation
parameter. Hence, the potential in the radial Hamiltonian (\ref{Hr})
is
\begin{eqnarray}\label{eq_m}
V_1= -\frac{\kappa}{r}+\beta r^l(1+\lambda
r^2)-\frac{1}{2}[(\frac{1}{2}-m^2)\lambda-\frac{m^2-1/4}{r^2}].
\end{eqnarray}
The hypervirial relation Eq. (\ref{eq:two-dimension diui}) turns to
\begin{eqnarray}
&&\frac{1}{4}\left[k(k-1)(k-2)-(k-1)(4m^2-1)\right]\langle r^{k-3}\rangle_\lambda+\frac{\lambda k}{2}(k^2+2-4m^2)\langle r^{k-1}\rangle_\lambda\nonumber\\
&&+2kE_{n}\langle r^{k-1}\rangle_\lambda+2(k-1)\kappa\langle r^{k-2}\rangle_\lambda+\frac{k}{4}(k+1)(k+2)\langle r^{k+1}\rangle_\lambda\nonumber\\
&&-\beta(2k+l)\langle
r^{k+l-1}\rangle_\lambda-\beta\lambda(2k+l+2)\langle
r^{k+l+1}\rangle_\lambda=0.
\end{eqnarray}
Considering the angular quantum  number $m^2$ as a parameter of the
potential $V_1$, one can obtain the  expansion  coefficients for
$\langle r^{-2}\rangle_\lambda$ by using the HF theorem,
\begin{eqnarray}
\langle r^{-2}\rangle_\lambda&=&\langle r^{-2}\rangle+\lambda\langle
1\rangle=2\frac{\partial E_{n}}{\partial m^2}.
\end{eqnarray}
From this starting point, as we show in the one-dimensional case, we
can get any order perturbation on the energy level, with the
precondition that $l$ is a negative integer.

Taking $l=-3$ for example, in the first approximation, the
eigenvalue $E_n$ is
\begin{eqnarray}
\hspace{-1in}E_{n}&=&-\frac{\kappa^2}{2(n+\sqrt{m^2}+\frac{1}{2})^2}+\frac{\lambda}{2}(n+\sqrt{m^2})(n+\sqrt{m^2}+\frac{1}{2})\nonumber\\
\hspace{-1in}&&+\beta\frac{8\kappa^3}{\sqrt{m^2}(4m^2-1)(n+\sqrt{m^2}+\frac{1}{2})^3}+\beta\frac{2\kappa\lambda}{\sqrt{m^2}(4m^2-1)}(4n+4\sqrt{m^2}+1).
\end{eqnarray}
When $\lambda\rightarrow0$, this result is coincided with the
literature \cite{mcrae1992canonical}.

\section {Conclusion and Discussion}\label{concl}

The VT in a spherical geometry has been proved in both classical and
quantum conditions. We also have considered the HVT and got the
hypervirial relations. The HVT and HF theorems have been shown to
provide a powerful method of generating perturbation expansions. We
have taken the Coulomb problem and harmonic oscillator for instances
to illustrate this method. When the curvature $\lambda$ is zero, the
results reduce to the counterpart of Euclidean space.

In this paper, we only give attention to one- and two-dimensional
systems. Since the Higgs' results have extended to the
$N$-dimensional spherical geometry directly
\cite{leemon1979dynamical}, we can foretell our treatment can be
generalized to the $N$-sphere and suggest the VT is given by
$\langle(1+\lambda r^2)\frac{\pi^2}{2}+\frac{\pi^2}{2}(1+\lambda
r^2)\rangle+\frac{1}{2}\langle\lambda(1+\lambda r^2)(N+3\lambda
r^2)\rangle=n\langle(1+\lambda r^2)\vec{r}\cdot \nabla V\rangle$.
Some researchers have discussed the superintegrable potentials in
the the hyperbolic plane \cite{hyperbolic}, it is interesting and
possible to study the VT , HVT and HVHF in the situation of the
curvature $\lambda<0$. On the other hand, the systems in the curved
space we investigate in this work also can be considered as the
problems with position-dependent effective mass, which  are widely
applied in various areas of material science and condensed matter
\cite{plastino1999supersymmetric,quesne2007spectrum,Bastard1988,serra1997spin}.
We hope to find the applications of our results in these directions
in the further research.

\section*{Acknowledgments}
We thank Lei Fang and Ci Song for their valuable discussions. This
work is supported by NSF of China (Grant No. 10975075) and the
Fundamental Research Funds for the Central Universities.

\section*{References}
\bibliography{viral}

\begin{thebibliography}{10}

\bibitem{clausius1870xvi}
R.~Clausius.
\newblock {XVI. On a mechanical theorem applicable to heat}.
\newblock {\em Philosophical Magazine Series 4}, 40(265):122--127, 1870.

\bibitem{born1926quantenmechanik}
M.~Born, W.~Heisenberg, and P.~Jordan.
\newblock {Zur Quantenmechanik. II.}
\newblock {\em Zeitschrift f\"{u}r Physik}, 35(8):557--615, 1926.

\bibitem{schiff1968quantum}
L.I. Schiff.
\newblock {\em {Quantum mechanics 3rd ed}}.
\newblock McGraw-Hill, 1968.

\bibitem{hirschfelder1960classical}
J.O. Hirschfelder.
\newblock {Classical and quantum mechanical hypervirial theorems}.
\newblock {\em The Journal of Chemical Physics}, 33:1462, 1960.

\bibitem{hellmann1937einf¨¹hrung}
H.~Hellmann.
\newblock {\em {Einf\"{u}hrung in die Quantenchemie}}.
\newblock Franz Deuticke, Vienna, 1937.

\bibitem{feynman1939forces}
RP~Feynman.
\newblock {Forces in molecules}.
\newblock {\em Physical Review}, 56(4):340--343, 1939.

\bibitem{swenson1972hypervirial}
R.J. Swenson and S.H. Danforth.
\newblock {Hypervirial and Hellmann-Feynman Theorems Applied to Anharmonic
  Oscillators}.
\newblock {\em The Journal of Chemical Physics}, 57:1734, 1972.

\bibitem{killingbeck1978perturbation}
J.~Killingbeck.
\newblock {Perturbation theory without wavefunctions}.
\newblock {\em Physics Letters A}, 65(2):87--88, 1978.

\bibitem{higgs1979dynamical}
P.~W. Higgs.
\newblock {Dynamical symmetries in a spherical geometry. I}.
\newblock {\em Journal of Physics A: Mathematical and General}, 12:309, 1979.

\bibitem{Bertrand}
Joseph Louis~Fran\c{c}ois Bertrand.
\newblock {Th{\'e}or\`{e}me relatif au mouvement d'un point attir{\'e} vers un
  centre fixe}.
\newblock {\em C. R. Acad. Sci.}, 77:849--853, 1873.

\bibitem{BertrandEnglish}
F~C Santos, V.~Soares, and A~C Tort.
\newblock {An English translation of Bertrand's theorem}.
\newblock {\em Arxiv preprint:0704.2396}, 2007.

\bibitem{bacry1966dynamical}
H.~Bacry, H.~Ruegg, and J.M. Souriau.
\newblock {Dynamical groups and spherical potentials in classical mechanics}.
\newblock {\em Communications in Mathematical Physics}, 3(5):323--333, 1966.

\bibitem{KK}
V~P Karassiov and A~B Klimov.
\newblock {An algebraic approach for solving evolution problems in some
  nonlinear quantum models}.
\newblock {\em Physics Letters. A}, 191(1-2):117--126, 1994.

\bibitem{zhang2009higgs}
Fu-Lin Zhang, Bo~Fu, and Jing-Ling Chen.
\newblock {Higgs algebraic symmetry in the two-dimensional Dirac equation}.
\newblock {\em Physical Review A}, 80(5):54102, 2009.

\bibitem{JLChenHiggs}
J.-L. Chen, Y.~Liu, and M.-L. Ge.
\newblock {Application of nonlinear deformation algebra to a physical system
  with P{\"{o}}schl-Teller potential}.
\newblock {\em Journal of physics A: mathematical and general}, 31:6473--6481,
  1998.

\bibitem{CS}
R.~Floreanini, L.~Lapointe, and L.~Vinet.
\newblock {The polynomial SU (2) symmetry algebra of the two-body Calogero
  model}.
\newblock {\em Physics Letters B}, 389(2):327--333, 1996.

\bibitem{PhysRevB.83.115129}
Martin Greiter.
\newblock Landau level quantization on the sphere.
\newblock {\em Phys. Rev. B}, 83(11):115129, 2011.

\bibitem{PhysRevLett.51.605}
F.~D.~M. Haldane.
\newblock Fractional quantization of the hall effect: A hierarchy of
  incompressible quantum fluid states.
\newblock {\em Phys. Rev. Lett.}, 51(7):605--608, 1983.

\bibitem{XJZ3}
Yi-Bing Ding.
\newblock In J.~Y. Zeng, G.~L. Long, and S.~Y. Pei, editors, {\em Recent
  Progress in Quantum Mechanics (Third Volume)}, page 286. Beijing: Tsinghua
  University, 2003.

\bibitem{de1991operator}
OL~De~Lange and RE~Raab.
\newblock {\em {Operator methods in quantum mechanics}}.
\newblock Oxford University Press, USA, 1991.

\bibitem{quesne2007spectrum}
C.~Quesne.
\newblock {Spectrum generating algebras for position-dependent mass oscillator
  Schr{\\"o}dinger equations}.
\newblock {\em Journal of Physics A: Mathematical and Theoretical}, 40:13107,
  2007.

\bibitem{mcrae1992canonical}
SM~McRae and ER~Vrscay.
\newblock {Canonical perturbation expansions to large order from classical
  hypervirial and Hellmann--Feynman theorems}.
\newblock {\em Journal of Mathematical Physics}, 33:3004, 1992.

\bibitem{leemon1979dynamical}
H.~I. Leemon.
\newblock {Dynamical symmetries in a spherical geometry. II}.
\newblock {\em Journal of Physics A: Mathematical and General}, 12:489, 1979.

\bibitem{hyperbolic}
M.F. Ra{\~n}ada and M.~Santander.
\newblock {Superintegrable systems on the two-dimensional sphere S and the
  hyperbolic plane H}.
\newblock {\em Journal of Mathematical Physics}, 40:5026, 1999.

\bibitem{plastino1999supersymmetric}
A.~R. Plastino, A.~Rigo, M.~Casas, F.~Garcias, and A.~Plastino.
\newblock {Supersymmetric approach to quantum systems with position-dependent
  effective mass}.
\newblock {\em Physical Review A}, 60(6):4318--4325, 1999.

\bibitem{Bastard1988}
G.~Bastard.
\newblock {\em {Wave Mechanics Applied to Semiconductor Heterostructure}}.
\newblock Les Editions de Physique, Les Ulis, France, 1988.

\bibitem{serra1997spin}
L.~Serra and E.~Lipparini.
\newblock {Spin response of unpolarized quantum dots}.
\newblock {\em Europhysics Letters}, 40:667, 1997.

\end{thebibliography}
\vspace{3cm}

\appendix

\section{Proof of the Virial Theorem in Classical Mechanics}\label{app}
In this part, we will give the strict proof for the classical VT in
Eqs. (\ref{02}) and (\ref{jingdian}). We adopt the subscripts $p$
and $s$ to distinguish the systems on a plane and on a sphere
respectively.
%
 From the Eq. (\ref{guiji}), we know that, for a given
$V(r)$, when
\begin{eqnarray}\label{ELSP}
E_s-\frac{1}{2}\lambda L_s^2=E_p, \ \ \ L_s=L_p,
\end{eqnarray}
the projected orbit of a spherical system is the same as the orbit
of a system in Euclidean geometry. It is easy to find that, for the
corresponding points $(r_s,\theta_s)=(r_p,\theta_p)=(r,\theta)$, the
velocities satisfy
\begin{eqnarray}\label{r}
\vec{v}_s=(1+\lambda r^2)\vec{v}_p,
\end{eqnarray}
where $\vec{v}_s=(\dot{r}_s,r_s \dot{\theta}_s)$ and
$\vec{v}_p=(\dot{r}_p,r_p \dot{\theta}_p)$. For the system in a flat
space whose Hamiltonian is given by $H=p^2/2+V$, the two terms in
Eq. (\ref{0}) are
\begin{eqnarray}\label{03}
&&\langle \vec{r}_p\cdot \nabla
V\rangle=\frac{1}{\tau_p}\int_0^{\tau_p} \vec{r}_p\cdot \nabla
Vdt_p=\frac{1}{\tau_p}\int_c\vec{r}_p\cdot \nabla V
\frac{1}{v_p^2}\vec{v_p}\cdot
d\vec{s}_p, \nonumber\\
&&\langle T_p\rangle=\frac{1}{\tau_p}\int_0^{\tau_p}
\frac{1}{2}v_p^2dt_p=\frac{1}{\tau_p}\int_c \frac{1}{2}v_p^2
\frac{1}{v_p^2}\vec{v_p}\cdot d\vec{s}_p,
\end{eqnarray}
where $d\vec{s}_p= (d r_p, r_p d \theta_p)$, $c$ denotes the orbit
of motion, and $\tau_p$ is the period (for the aperiodic case
${\tau_p}\rightarrow +\infty$). Suppose the period of the system
with the same orbit $c$ in the sphere described by Eq.
(\ref{Hamilton}) is $\tau_s$. Then, considering the relations in
Eqs. (\ref{ELSP}) and (\ref{r}), one can find
\begin{eqnarray}
&&\langle \vec{r}_p\cdot \nabla
V\rangle=\frac{\tau_s}{\tau_p}\langle
 (1+\lambda r_s^2)\vec{r}_s\cdot\nabla
V\rangle, \nonumber\\
&&\langle T_p\rangle=\frac{\tau_s}{\tau_p}\big[ \langle(1+\lambda
r_s^2){T_{s}}_r\rangle+\langle {T_{s}}_\theta\rangle\big]
=\frac{\tau_s}{\tau_p}\langle(1+\lambda
r_s^2)\frac{\pi_s^2}{2}\rangle,
\end{eqnarray}
where the radial kinetic energy
${T_{s}}_r=R^2\dot{\chi}_s^2/2=\dot{r}_s^2/[2(1+\lambda r_s^2)^2]$
and the rotational kinetic energy ${T_{s}}_\theta=R^2 \sin^2 \chi_s
\dot{\theta}_s^2/2=r_s^2 \dot{\theta}_2^2/[2(1+\lambda r_s^2)]$.
Therefore, the relation in Eq. (\ref{02}) is the VT in a spherical
geometry, and it equivalents to Eq. (\ref{jingdian}). Here the proof
comes to an end.

\end{document}